\documentclass[aps,prd,twocolumn,nofootinbib,longbibliography,10pt]{revtex4-2}
\usepackage{amsmath,amssymb,bbold}
\usepackage{graphicx,epsfig}
\RequirePackage{epstopdf}
\graphicspath{}
\usepackage[colorlinks=true,urlcolor=blue,citecolor=red,linkcolor=blue]{hyperref}
\begin{document}
	\title{Density matrix analysis of systems with periodic Hamiltonians}
	\author{Soham Sen}
	\email{sensohomhary@gmail.com}
	\affiliation{Department of Astrophysics and High Energy Physics, S. N. Bose National Centre for Basic Sciences, JD Block, Sector-III, Salt Lake City, Kolkata-700 106, India}
	\author{Manjari Dutta}
	\email{chandromouli15@gmail.com}
	\affiliation{Department of Astrophysics and High Energy Physics, S. N. Bose National Centre for Basic Sciences, JD Block, Sector-III, Salt Lake City, Kolkata-700 106, India}
		\author{Sunandan Gangopadhyay}
	\email{sunandan.gangopadhyay@gmail.com}
	\affiliation{Department of Astrophysics and High Energy Physics, S. N. Bose National Centre for Basic Sciences, JD Block, Sector-III, Salt Lake City, Kolkata-700 106, India}

	\begin{abstract}
	\noindent
	In this work, we consider simple systems that are governed by Hamiltonians with time periodicity. Our analysis is mainly focused on the density matrix approach and aims to solve the von Neumann equation of motion from which one can extract the state of the system when the system is in a pure state. We start our analysis with the standard Rabi-oscillation problem. We consider a density matrix corresponding to the entire model system and solve the von Neumann equation of motion. We have then made use of the Lewis-Reisenfeld invariant approach and arrived at the exact same result, which implies that the density matrix of the system can indeed be identified with the Lewis invariant. Then we consider a two-level system with a constant magnetic field in the $z$-direction and a time-dependent magnetic field in the $x$-direction. We solve the von Neumann equation of motion for this system and calculate the various coherence measures, and plot them to investigate the time dependence and reliability of different coherence measures.
	\end{abstract}
	\maketitle
	\section{Introduction}\label{S1}
\noindent From the advent of quantum mechanics at the beginning of the twentieth century, the investigation of systems governed by time-dependent Hamiltonians is of great interest as it very closely portrays real physical systems. In quantum mechanics, a system is represented by a state that evolves through time via the influence of a Hamiltonian. The most general way to obtain a state of the system is by solving the time-dependent Schr\"{o}dinger's equation of motion. While dealing with state vectors for a system, it is important to remember that for a multi-level system, the system can also be in a mixed state, such that one cannot truly write down a state vector of the system. Hence, the way out is to write down a density matrix and solve the von Neumann equation of motion. Even for systems with a pure state, the density matrix of the system gives a better perspective from a dynamical as well as information theoretical point of view. The reason for this is that it helps one to write down several coherence measures that portray different quantum mechanical effects while considering an open quantum system. The problem with such systems influenced by a time-dependent Hamiltonian lies in the fact that very few models are exactly solvable \cite{Landau,Zener,Rabi,BarnsSarma}. The Rabi oscillation problem \cite{Rabi} is one of the more interesting exactly solvable models. Rabi oscillations can be found in Rydberg atoms \cite{Rydberg}, quantum information science, optical spectra \cite{Atomic1}, atomic physics \cite{Atomic2}, Josephson junctions \cite{Josephson,Josephson1,Josephson2}, cavity quantum mechanics \cite{Cavity1,Cavity2,Cavity3,Cavity4,Cavity5} and many other physical as well as experimental scenarios. In \cite{AJPMerlin}, some exactly solvable models regarding the Rabi-oscillation problems has been revisited. 

\noindent These Rabi oscillations arise due to the influence of a periodically time-varying field on multi-level quantum systems. For a quantum system in the presence of such a periodic time-varying field, the stationary solution gives a unique kind of states which are also known as the Floquet states. Such states also bear the same periodicity of the system Hamiltonian. Floquet's theorem \cite{Floquet1,Floquet2} states that in the presence of such a periodic Hamiltonian, one can find a complete set of solutions corresponding to the system which has the general structure of a phase term multiplied by the periodic Floquet states. Such states are of high interest as the phases can unveil underlying topological properties and have been intensely investigated in the literature \cite{FloquetState0,FloquetState1,
FloquetState2,FloquetState3,FloquetState4}. 

\noindent A very unique way of dealing with quantum systems driven by time-dependent Hamiltonians lies in the method of finding out an invariant operator that has explicitly been showcased in \cite{HRLewis1,HRLewis2}. The invariant is obtained by keeping in consideration that the total time derivative of the invariant must vanish. One can then find out such invariant operators which later on help one to find out the time-dependent effective phase corresponding to the model system. The invariant operator is also known as the Lewis invariant, and the phase is termed the Lewis phase.  

\noindent In this work, we primarily consider two-level systems being driven by a periodically time-varying field, and we have made use of the density matrix approach to solve such models. At first, we have investigated the Rabi-oscillation problem (under the rotating-wave approximation) and solved the corresponding von-Neumann equations of motion. We have then investigated the model using the Lewis-invariant approach, and our aim there is to find a connection between the Lewis-invariant approach and the standard density matrix approach. Finally, we have considered a two-level atom which is being driven by a magnetic field of which the $x$-component is influenced by a periodic-square pulse. Now, working with density matrices also lets one ask the question whether the coherence remains intact when driven by such periodic Hamiltonians. The way to investigate the coherence is to calculate several coherence measures using the density matrix of the model system. At first, we have considered the $l_1$-norm measure of coherence that is very intuitive and is defined as \cite{CoherenceMeasure}
\begin{equation}\label{1.1}
C_{l_1}(\rho)\equiv\sum\limits_{\substack{{i,j=1}\\{d}}}|\rho_{ij}|
\end{equation}
where the absolute values of all the off-diagonal components of the density matrix are summed over, and $d$ denotes the dimension of the Hilbert space. Another way to measure coherence is by calculating the Frobenius-norm measure of coherence that is defined as \cite{FrobeniusMeasure}
\begin{equation}\label{1.2}
C_F(\rho)\equiv \sqrt{\frac{d}{d-1}}||\rho-\rho_{\mathbb{1}}||_F
\end{equation}
with $\rho_{\mathbb{1}}=\frac{\mathbb{1}_d}{d}$ denoting the maximally mixed states. The Frobenius-norm measure of coherence is invariant under unitary transformation and is basis-independent. The coherence measure considers the geometric distance between the state $\rho$ from the maximally mixed state $\rho_{\mathbb{1}}$, which indicates that the Frobenius norm measure of coherence has a very clear geometric interpretation.

\noindent Our work is organized as follows. In section (\ref{S2}), we consider the Rabi-oscillation problem, which has been solved using the density matrix as well as the Lewis-invariant approach. In section (\ref{S3}), we consider a two-level system driven by a time-varying magnetic field and have calculated the two coherence measures discussed earlier. Finally, in section (\ref{S4}), we summarize our results.
\section{A different approach to solve the Rabi oscillation problem}\label{S2}
\noindent The basic model consists of a two-level atom with the ground state denoted by $|g\rangle$ with energy $E_g$ and the excited state denoted by $|e\rangle$ with energy $E_e$. A periodically varying classical field is applied with amplitude $\alpha_0$ and frequency $\omega_0$. The analytical form of the Hamiltonian reads
\begin{equation}\label{2.1}
\begin{split}
\hat{H}&=E_g|g\rangle\langle g|+E_e|e\rangle\langle e|+\alpha_0^*\Lambda_{eg}^*\cos \omega_0 t|g\rangle\langle e|\\&+\alpha_0\Lambda_{eg}\cos \omega_0 t|e\rangle\langle g|~.
\end{split}
\end{equation}
The standard way to exactly solve the Rabi problem is to consider the rotating wave approximation, which recasts the above Hamiltonian as
\begin{equation}\label{2.2}
\begin{split}
\hat{H}_{R}&=E_g|g\rangle\langle g|+E_e|e\rangle\langle e|+\alpha_0^*\Lambda_{ge}e^{i \omega_0 t}|g\rangle\langle e|\\&+\alpha_0\Lambda_{eg}e^{-i \omega_0 t}|e\rangle\langle g|~.
\end{split}
\end{equation}
Instead of solving the Schr\"{o}dinger's equation of motion, we start by writing down the initial density matrix for the system. Considering the system to be in a ground state, we can write down the initial system-density matrix as\footnote{As we shall be expressing the density matrix as well as the Hamiltonian in a matrix form, the operator notations have been removed.}
\begin{equation}\label{2.3}
\rho_i=\rho(t=0)=\begin{pmatrix}
1&&0\\
0&&0
\end{pmatrix}
\end{equation}
which is constructed by executing the outer product between the column matrix $|g\rangle=\begin{pmatrix}
1\\0
\end{pmatrix}$ with the row matrix $\langle g|=\begin{pmatrix}
1&&0
\end{pmatrix}$.
The Hamiltonian in eq.(\ref{2.2}) can be expressed in a matrix form as
\begin{equation}\label{2.4}
H_R(t)=
\begin{pmatrix}
E_g && \alpha_0^*\Lambda_{ge}e^{i\omega_0 t}\\
\alpha_0\Lambda_{eg}e^{-i\omega_0 t} && E_e
\end{pmatrix}
\end{equation}
where the Hermiticity condition of the Hamiltonian gives $\Lambda_{ge}=\Lambda_{eg}^*$.
The density matrix at any arbitrary time $t$ takes the form
\begin{equation}\label{2.5}
\rho(t)=\begin{pmatrix}
\rho_{gg}(t)&&\rho_{ge}(t)\\
\rho_{eg}(t)&&\rho_{ee}(t)
\end{pmatrix}~.
\end{equation}
At $t=0$, the above equation must reduce to the form of the density matrix in eq.(\ref{2.3}) and therefore, we have the set of initial conditions for the elements of the density matrix as $\rho_{gg}(0)=1$ and $\rho_{ge}(0)=\rho_ {eg}(0)=\rho_{ee}(0)=0$.
\subsection{von Neumann equation of motion}\label{IIA}
\noindent To obtain the analytical form of the density matrix, we need to solve the quantum Liouville equations given as
\begin{equation}\label{2.A.6}
\frac{\partial\hat{\rho}}{\partial t}=-\frac{i}{\hbar}[\hat{H}_R,\hat{\rho}]~.
\end{equation}
Using eq.(s)(\ref{2.4},\ref{2.5}), one can obtain the matrix form of $[\hat{H}_R,\hat{\rho}]$ and just taking a time derivative of eq.(\ref{2.5}), one can obtain the left hand side of the above equation. Comparing the two results and setting $\hbar=1$, one obtains four ordinary differential equations of motion given by
\begin{align}
\dot{\rho}_{gg}&=-i\left(\rho_{eg}\alpha_0^*\Lambda_{ge}e^{i\omega_0 t}-\rho_{ge}\alpha_0\Lambda_{eg}e^{-i\omega_0 t}\right)\label{2.A.7}\\
\dot{\rho}_{ee}&=i\left(\rho_{eg}\alpha_0^*\Lambda_{ge}e^{i\omega_0 t}-\rho_{ge}\alpha_0\Lambda_{eg}e^{-i\omega_0 t}\right)\label{2.A.8}\\
\dot{\rho}_{ge}&=-i\left(\rho_{ge}(E_g-E_e)+(\rho_{ee}-\rho_{gg})\alpha_0^*\Lambda_{ge}e^{i\omega_0 t}\right)\label{2.A.9}\\
\dot{\rho}_{eg}&=i\left(\rho_{eg}(E_g-E_e)+(\rho_{ee}-\rho_{gg})\alpha_0\Lambda_{eg}e^{-i\omega_0 t}\right)\label{2.A.10}~.
\end{align}
Our aim is to obtain the time dependence of all the elements of the density matrix. At first, we add eq.(\ref{2.A.7}) with eq.(\ref{2.A.8}) which gives the new equation
\begin{equation}\label{2.A.11}
\dot{\rho}_{gg}(t)+\dot{\rho}_{ee}(t)=0~.
\end{equation}
Executing the first order time integral in the above equation, we arrive at the equation given by
\begin{equation}\label{2.A.12}
\rho_{gg}(t)=-\rho_{ee}(t)+a_0
\end{equation}
where $a_0$ is an integration constant. Now, for a density matrix describing a system, the trace must be unity. Hence, $\text{tr}(\rho)=\rho_{gg}(t)+\rho_{ee}(t)=1$ at any arbitrary time $t$. Using the above result back in eq.(\ref{2.A.12}), we obtain the value of the constant to be $a_0=1$. For proceeding further, we multiply eq.(\ref{2.A.9}) with $\alpha_0 \Lambda_{eg}e^{-i\omega_0t}$ and multiplying eq.(\ref{2.A.10}) with $\alpha_0^* \Lambda_{ge}e^{i\omega_0t}$, we arrive at two new equations given by
\begin{align}
\alpha_0\Lambda_{eg}e^{-i\omega_0 t}\dot{\rho}_{ge}(t)=& -i\alpha_0\Lambda_{eg}e^{-i\omega_0 t}\rho_{ge}(t)(E_g-E_e)\nonumber\\&-i|\alpha_0\Lambda_{eg}|^2(\rho_{ee}(t)-\rho_{gg}(t))\label{2.A.13}\\
\alpha_0^*\Lambda_{ge}e^{i\omega_0 t}\dot{\rho}_{eg}(t)=& i\alpha_0^*\Lambda_{ge}e^{i\omega_0 t}\rho_{eg}(t)(E_g-E_e)\nonumber\\&+i|\alpha_0\Lambda_{eg}|^2(\rho_{ee}(t)-\rho_{gg}(t))\label{2.A.14}~.
\end{align}
Adding eq.(\ref{2.A.13}) with eq.(\ref{2.A.14}), we arrive at the following new equation given as
\begin{equation}\label{2.A.15}
\begin{split}
&\alpha_0\Lambda_{eg}e^{-i\omega_0 t}\dot{\rho}_{ge}(t)+\alpha_0^*\Lambda_{ge}e^{i\omega_0 t}\dot{\rho}_{eg}(t)\\&=-i(E_g-E_e)\left(\rho_{ge}\alpha_0\Lambda_{eg}e^{-i\omega_0 t}-\rho_{eg}\alpha_0^*\Lambda_{eg}^*e^{i\omega_0 t}\right)~.
\end{split}
\end{equation}
Writing the left-hand side of the above equation as a total time derivative, it is possible to recast eq.(\ref{2.A.15}) as

\begin{equation}\label{2.A.16}
\begin{split}
&\frac{d}{dt}\left(\alpha_0\Lambda_{eg}e^{-i\omega_0 t}\rho_{ge}(t)+\alpha_0^*\Lambda_{ge}e^{i\omega_0 t}\rho_{eg}(t)\right)\\
&=i(E_e-E_g-\omega_0)\left(\rho_{ge}\alpha_0\Lambda_{eg}e^{-i\omega_0 t}-\rho_{eg}\alpha_0^*\Lambda_{eg}^*e^{i\omega_0 t}\right)\\
&=(E_e-E_g-\omega_0)\dot{\rho}_{gg}(t)
\end{split}
\end{equation}
where in the last line of the above equation, we have made use of eq.(\ref{2.A.7}). Executing the time integral on both sides of the above equations, we arrive at the relation
\begin{equation}\label{2.A.17}
\alpha_0\Lambda_{eg}e^{-i\omega_0 t}\rho_{ge}+\alpha_0^*\Lambda_{ge}e^{i\omega_0 t}\rho_{eg}=(E_e-E_g-\omega_0)\rho_{gg}+\alpha_1
\end{equation}
with $\alpha_1$ being an integration constant. To obtain the value of the integration constant, we need to use the initial condition corresponding to the elements of the density matrix. We already know that at $t=0$, $\rho_{gg}(0)=1$, and $\rho_{ge}(0)=\rho_{eg}(0)=0$. These initial values help us to set the value of the constant $\alpha_1$ to be
\begin{equation}\label{2.A.18}
\alpha_1=E_g-E_e+\omega_0.
\end{equation}
For expressional simplicity, we define a new constant $\Theta\equiv E_e-E_g-\omega_0$ \cite{AJPMerlin}. Next, we subtract eq.(\ref{2.A.14}) from eq.(\ref{2.A.13}) and, as before, expressing the left-hand side as a total time derivative, we arrive at a new equation given as
\begin{widetext}
\begin{equation}\label{2.A.19}
\begin{split}
\frac{d}{dt}\left(\alpha_0\Lambda_{eg}e^{-i\omega_0 t}\rho_{ge}(t)-\alpha_0^*\Lambda_{ge}e^{i\omega_0 t}\rho_{eg}(t)\right)&=i(E_e-E_g-\omega_0)\left(\alpha_0\Lambda_{eg}e^{-i\omega_0 t}\rho_{ge}(t)+\alpha_0^*\Lambda_{ge}e^{i\omega_0 t}\rho_{eg}(t)\right)\\&-2i|\alpha_0\Lambda_{eg}|^2\left(\rho_{ee}(t)-\rho_{gg}(t)\right)\\
\implies i\frac{d}{dt}\left(\alpha_0\Lambda_{eg}e^{-i\omega_0 t}\rho_{ge}(t)-\alpha_0^*\Lambda_{ge}e^{i\omega_0 t}\rho_{eg}(t)\right)&=-(E_e-E_g-\omega_0)\left(\alpha_0\Lambda_{eg}e^{-i\omega_0 t}\rho_{ge}(t)+\alpha_0^*\Lambda_{ge}e^{i\omega_0 t}\rho_{eg}(t)\right)\\&-2|\alpha_0\Lambda_{eg}|^2\left(2\rho_{gg}(t)-1\right)\\
&=-(E_e-E_g-\omega_0)^2\left(\rho_{gg}-1\right)-2|\alpha_0\Lambda_{eg}|^2\left(2\rho_{gg}(t)-1\right)
\end{split}
\end{equation} 
where in the last line of the above equation, we have made use of eq.(\ref{2.A.12}) and eq.(\ref{2.A.17}) with the forms of $\alpha_0$ and $\alpha_1$.
\end{widetext}
Making use of eq.(\ref{2.A.7}), we can write down a second-order differential equation in $\rho_{gg}$ from eq.(\ref{2.A.19}) as
\begin{equation}\label{2.A.20}
\ddot{\rho}_{gg}(t)+\left(\Theta^2+4|\alpha_0\Lambda_{eg}|^2\right)\rho_{gg}(t)=\Theta^2+2|\alpha_0\Lambda_{eg}|^2~.
\end{equation}
Defining a new constant $\Omega^2=\frac{\Theta^2}{4}+|\alpha_0\Lambda_{eg}|^2$, one can now obtain the analytical time dependent form of $\rho_{gg}(t)$ as
\begin{equation}\label{2.A.21}
\rho_{gg}(t)=\beta_1\cos 2\Omega t+\beta_2\sin 2\Omega t+\frac{\Theta^2+2|\alpha_0\Lambda_{eg}|^2}{4\Omega^2}
\end{equation}
with $\beta_1$ and $\beta_2$ being integration constants. From eq.(\ref{2.A.7}), it is easy to find out using the initial conditions for the density matrix that at $t=0$, $\dot{\rho}_{gg}(0)=0$. Now taking time derivative of both sides of eq.(\ref{2.A.21}) and using the condition $\dot{\rho}_{gg}=0$ at $t=0$, we obtain $\beta_2=0$.  At $t=0$, again $\rho_{gg}(0)=1$, which gives us the analytical form of the constant $\beta_1$ as $\beta_1=\frac{|\alpha_0\Lambda_{eg}|^2}{2\Omega^2}$. One can write down the solution for $\rho_{gg}(t)$ as
\begin{equation}\label{2.A.22}
\rho_{gg}(t)=\frac{|\alpha_0\Lambda_{eg}|^2}{2\Omega^2}\cos 2\Omega t+\frac{\Theta^2+2|\alpha_0\Lambda_{eg}|^2}{4\Omega^2}~.
\end{equation}
From eq.(\ref{2.A.12}) or the trace condition of the density matrix, it is then easy to obtain the analytical form of $\rho_{ee}(t)$, which is given as
\begin{equation}\label{2.A.23}
\begin{split}
\rho_{ee}(t)&=1-\rho_{gg}(t)\\&=\frac{|\alpha_0\Lambda_{eg}|^2}{2\Omega^2}\left(1-\cos2\Omega t\right)\\
&=\frac{|\alpha_0\Lambda_{eg}|^2}{\Omega^2}\sin^2\Omega t~.
\end{split}
\end{equation}
Now, substituting the analytical forms of $\rho_{gg}(t)$ and $\rho_{ee}(t)$ from eq.(s)(\ref{2.A.22},\ref{2.A.23}) in eq.(\ref{2.A.9}), we obtain the ordinary differential equation in $\rho_{ge}(t)$ as
\begin{equation}\label{2.A.24}
\begin{split}
\dot{\rho}_{ge}(t)&=i\rho_{ge}(t)(E_e-E_g)+\left(1-\frac{|\alpha_0\Lambda_{eg}|^2}{\Omega^2}\right)i\alpha_0 \Lambda_{ge} e^{i\omega_0 t}\\
&+\frac{i|\alpha_0\Lambda_{eg}|^2}{\Omega^2}\alpha_0^*\Lambda_{ge}e^{i\omega_0 t}\cos2 \Omega t.
\end{split}
\end{equation}
The solution of the above differential equation, after a little bit of simplification, reads
\begin{equation}\label{2.A.25}
\begin{split}
\rho_{ge}(t)&=\frac{\alpha_0^*\Lambda_{ge}e^{i\omega_0 t}}{4\Omega^2}\left(2 i\Omega \sin 2\Omega t+\Theta \cos 2\Omega t-\Theta \right)\\
&+\beta_3 e^{i(E_e-E_g)t}
\end{split}
\end{equation}
with $\beta_3$ being an integration constant. At $t=0$, $\rho_{ge}(0)=0$ then $\beta_3=0$. The analytical form of $\rho_{ge}$ then reads
\begin{equation}\label{2.A.26}
\rho_{ge}(t)=\frac{\alpha_0^*\Lambda_{ge}e^{i\omega_0 t}}{4\Omega^2}\left(\Theta \cos 2\Omega t-\Theta +2 i\Omega \sin 2\Omega t\right)~.
\end{equation}
Now $\rho_{eg}$ is just complex conjugate of $\rho_{ge}$ and therefore the analytical form of $\rho_{eg}$ reads
\begin{equation}\label{2.A.27}
\rho_{eg}(t)=\frac{\alpha_0\Lambda_{eg}e^{-i\omega_0 t}}{4\Omega^2}\left(\Theta \cos 2\Omega t-\Theta-2 i\Omega \sin 2\Omega t \right)~.
\end{equation}
Combining eq.(s)(\ref{2.A.22},\ref{2.A.23},\ref{2.A.26},\ref{2.A.27}) and after a little bit of simplification, one can write down the complete time-dependent form of the density matrix as
\begin{widetext}
\begin{equation}\label{2.A.28}
\rho(t)=\begin{pmatrix}
\cos^2\Omega t+\frac{\Theta^2}{4\Omega^2}\sin^2\Omega t&&\frac{\alpha_0^*\Lambda_{ge}e^{i\omega_0 t}}{4\Omega^2}\left(\Theta \cos 2\Omega t-\Theta +2 i\Omega \sin 2\Omega t\right)\\
\frac{\alpha_0\Lambda_{eg}e^{-i\omega_0 t}}{4\Omega^2}\left(\Theta \cos 2\Omega t-\Theta-2 i\Omega \sin 2\Omega t \right)&&\frac{|\alpha_0\Lambda_{eg}|^2}{\Omega^2}\sin^2\Omega t
\end{pmatrix}~.
\end{equation}
\end{widetext}
It is easy to check from the above form of the density matrix that $\rho^2=\rho$, indicating that the state remains pure at all times. Hence, one can obtain a time dependent pure state $|\phi(t)\rangle$ such that $\hat{\rho}(t)=|\phi(t)\rangle\langle \phi(t)|$. Our starting ansatz is 
\begin{equation}\label{2.A.29}
|\phi(t)\rangle=\mathcal{A}_1(t)|g\rangle+e^{i\kappa t}\mathcal{A}_2(t)|e\rangle
\end{equation}
where we need to obtain the analytical forms of $\mathcal{A}_1(t)$ and $\mathcal{A}_2(t)$. In the above equation, $\kappa$ is the relative phase between the two states of the system at any time $t$. If we execute the operation $|\phi(t)\rangle\langle\phi(t)|$, and compare it with the $\{g,g\}$ element of the density matreix in eq.(\ref{2.A.28}), we obtain
\begin{equation}\label{2.A.30}
\mathcal{A}_1(t)\mathcal{A}_1^*(t)=\cos^2\Omega t+\frac{\Theta^2}{4\Omega^2}\sin^2\Omega t.
\end{equation}
This gives the analytical form of $\mathcal{A}_1(t)$ to have two possibilities
\begin{equation}\label{2.A.31}
\mathcal{A}_1^{\pm}(t)=\cos\Omega t\pm\frac{i\Theta}{2\Omega}\sin\Omega t~.
\end{equation}
Next, we compare the coefficient of the $|e\rangle\langle e|$ component with the $\{e,e\}$ element of the density matrix in eq.(\ref{2.A.28}), which gives us 
\begin{equation}\label{2.A.32}
\mathcal{A}_2(t)\mathcal{A}_2^*(t)=\frac{|\alpha_0\Lambda_{eg}|^2}{\Omega^2}\sin^2\Omega t.
\end{equation}
Now, we are stuck with four choices for $\mathcal{A}_2(t)$ as
\begin{equation}\label{2.A.33}
\mathcal{A}_2^{\pm}(t)=\pm \frac{i \alpha_0\Lambda_{eg}}{\Omega}\sin\Omega t,~{\mathcal{A}'}_2^{\pm}(t)=\pm \frac{i \alpha_0^*\Lambda_{ge}}{\Omega}\sin\Omega t~.
\end{equation}
Combining eq.(\ref{2.A.31}) with eq.(\ref{2.A.33}) in eq.(\ref{2.A.29}), we can find the coefficient corresponding to the $|g\rangle\langle e|$ as well as the $|e\rangle\langle g|$ element. Comparing these coefficients with the off-diagonal components of the density matrix in eq.(\ref{2.A.28}), one obtains $\kappa=-\omega_0(t)$, $\mathcal{A}_1(t)=\mathcal{A}_1^+(t)$, and $\mathcal{A}_2(t)=\mathcal{A}_2^{-}(t)$. Hence, one obtains the form of the state in eq.(\ref{2.A.29}) as
\begin{equation}\label{2.A.34}
|\phi\rangle=\left[\cos\Omega t+\frac{i\Theta}{2\Omega}\sin\Omega t\right]|g\rangle-\frac{i\alpha_0\Lambda_{eg}}{\Omega}e^{-i\omega_0 t}\sin\Omega t|e\rangle.
\end{equation}
The above state is not an eigenstate of the Hamiltonian in eq.(\ref{2.4}). To find the set of eigenstates, one needs to make use of Floquet's theorem \cite{Floquet1,Floquet2}. As the Hamiltonian is periodic in time, the Floquet theorem states that there is a complete set of solutions of the form $e^{i\zeta t}|\phi(t)\rangle$ such that
\begin{equation}\label{2.A.35}
\langle\phi(t)|i\frac{\partial}{\partial t}-\hat{H}_R(t)|\phi(t)\rangle=\zeta~.
\end{equation}
Using eq.(\ref{2.A.34}), we obtain the analytical form of $\zeta$ to be 
\begin{equation}\label{2.A.36}
\zeta=\frac{1}{2}(\omega_0-E_e-E_g)~.
\end{equation}
Hence, we can write down the solution for the system as
\begin{equation}\label{2.A.37}
\begin{split}
|\psi(t)\rangle=&e^{\frac{i(\omega_0-E_e-E_g)t}{2}}|\phi(t)\rangle\\
=&e^{\frac{i(\omega_0-E_e-E_g)t}{2}}\Biggr[\left(\cos\Omega t+\frac{i\Theta}{2\Omega}\sin\Omega t\right)|g\rangle\\&-\frac{i\alpha_0\Lambda_{eg}}{\Omega}e^{-i\omega_0 t}\sin\Omega t|e\rangle\Biggr].
\end{split}
\end{equation}
In the next section, we shall approach solving the same model in the Lewis-Reisenfeld invariant approach.
\subsection{Lewis-Reisenfeld invariant approach}\label{IIB}
\noindent The Lewis-Riesenfeld invariant approach introduced in \cite{LewisRiesenfeld}, relies on finding out a Hermitian operator $\hat{I}(t)$ such that 
\begin{equation}\label{2.B.38}
\frac{d\hat{I}(t)}{dt}=\frac{\partial\hat{I}(t)}{\partial t}+\frac{1}{i}[\hat{I}(t),\hat{H}_R(t)]=0~.
\end{equation}
Following the form of the Hamiltonian in eq.(\ref{2.2}), it is possible to write down the most general form of the invariant operator as
\begin{equation}\label{2.B.39}
\hat{I}(t)=\delta_1(t)|g\rangle\langle g|+\gamma_{1}(t)|g\rangle\langle e|+\gamma_{2}(t)|e\rangle\langle g|+\delta_{2}(t)|e\rangle\langle e|
\end{equation}
with $\delta_1$, $\delta_2$, $\gamma_1$, and $\gamma_2$ are unknown time dependent coefficients. For a time dependent state vector $|\psi_{LR}(t)\rangle$, it satisfies the time-dependent Schr\"{o}dinger equation $i\frac{\partial |\psi_{LR}(t)\rangle}{\partial t}=\hat{H}_R|\psi_{LR}(t)\rangle$, then from eq.(\ref{2.B.38}), it is also straightforward to find that $\hat{I}(t)|\psi_{LR}(t)\rangle$ also satisfies the same Schr\"{o}dinger's equation of motion. Suppose that $\hat{I}(t)$ creates a complete set of commuting observables, and as a result, one can find a complete set of eigenstates of the Hermitian operator $\hat{I}(t)$. If $|\phi_{LR}(t)\rangle$ denotes the eigenstate of the invariant operator, then $\hat{I}(t)|\phi_{LR}(t)\rangle=\lambda |\phi_{LR}(t)\rangle$ with $\lambda$ being the eigenvalue. This eigenstate $ |\phi_{LR}(t)\rangle$ is related with the state vector $ |\psi_{LR}(t)\rangle$ via the relation
\begin{equation}\label{2.B.40}
  |\psi_{LR}(t)\rangle=e^{i\theta(t)}|\phi_{LR}(t)\rangle
\end{equation}
where $\theta(t)$ is also known as the Lewis phase. For analytical simplicity, we can define a new invariant operator $\hat{\mathcal{I}}(t)=\frac{\hat{I}(t)}{\lambda}$. The new operator holds the same structure as the ansatz in eq.(\ref{2.B.39}) provided the coefficients are scaled by a factor of $\frac{1}{\lambda}$ as $\delta_1^\lambda(t)\equiv \frac{\delta_1(t)}{\lambda}$, $\delta_2^\lambda(t)\equiv \frac{\delta_2(t)}{\lambda}$, $\gamma_1^\lambda(t)\equiv \frac{\gamma_1(t)}{\lambda}$, and $\gamma_2^\lambda(t)\equiv \frac{\gamma_2(t)}{\lambda}$. This new rescaled invariant operator also satisfies eq.(\ref{2.B.38}). 
Using the ansatz of the invariant operator in eq.(\ref{2.B.39}) and the form of the Hamiltonian from eq.(\ref{2.2}), back in eq.(\ref{2.B.38}), we arrive at four ordinary differential equations of the form
\begin{align}
i\dot{\delta}_1^\lambda&=\gamma_2^\lambda\alpha_0^*\Lambda_{ge}e^{i\omega_0 t}-\gamma_1^\lambda\alpha_0\Lambda_{eg}e^{-i\omega_0 t}\label{2.B.41}\\
i\dot{\delta}_2^\lambda&=-\gamma_2^\lambda\alpha_0^*\Lambda_{ge}e^{i\omega_0 t}+\gamma_1^\lambda\alpha_0\Lambda_{eg}e^{-i\omega_0 t}\label{2.B.42}\\
i\dot{\gamma}_1^\lambda&=\gamma_1^\lambda(E_g-E_e)+(\delta_2^\lambda-\delta_1^\lambda)\alpha_0^*\Lambda_{ge}e^{i\omega_0 t}\label{2.B.43}\\
i\dot{\gamma}_{2}^\lambda&=-\gamma_2^\lambda(E_g-E_e)-(\delta_2^\lambda-\delta_1 ^\lambda)\alpha_0\Lambda_{eg}e^{-i\omega_0 t}\label{2.B.44}~.
\end{align}
Taking $\delta_1^\lambda(t)=\xi^2(t)$, and adding eq.(s)(\ref{2.B.41},\ref{2.B.42}), we obtain $\dot{\delta}_2^\lambda(t)=-2\xi\dot{\xi}$. This implies $\delta_2^\lambda(t)=C-\xi^2(t)$ with $C$ being an integration constant. At $t=0$, $|\psi_{LR}(0)\rangle=|g\rangle$ as the system is in the ground state and as a result $\theta(0)=0$ with $|\phi_{LR}(0)\rangle=|g\rangle$ as well. Now, we know that $\hat{\mathcal{I}}(0)|\phi_{LR}(0)\rangle=|\phi_{LR}(0)\rangle$ at $t=0$. The standard action of $\hat{\mathcal{I}}(0)$ on $|\phi_{LR}(0)\rangle$ gives
\begin{equation}\label{2.B.45}
\begin{split}
\hat{\mathcal{I}}(0)|g\rangle&=\delta_1^\lambda(0)|g\rangle+\gamma_2^\lambda(0)|e\rangle~.
\end{split}
\end{equation}
The right-hand side must be equal to $|g\rangle$ following the eigenvalue equation of  $\hat{\mathcal{I}}(t)$. This gives $\gamma_2^\lambda(0)=0$ and $\delta_1^\lambda(0)=1$. This condition gives $\xi^2(0)=1$. As $\hat{\mathcal{I}}(t)$ is Hermitian, we can easily conclude $\gamma_1^\lambda(0)=0$. Correctly multiplying constants and combining the differential equations from eq.(s)(\ref{2.B.41},\ref{2.B.42},\ref{2.B.43},\ref{2.B.44}), we arrive at a differential equation in $\xi$ as
\begin{equation}\label{2.B.46}
\begin{split}
\frac{\ddot{\xi}}{\xi}+\frac{\dot{\xi}^2}{\xi^2}+4\Omega^2=\frac{1}{\xi^2}\left(\frac{\Theta^2}{2}+C|\alpha_0\Lambda_{eg}|^2\right)~.
\end{split}
\end{equation}
The above equation for the current-model system is well-known as the Ermakov-Pinney equation. One can obtain a standard solution after a little bit of rearrangement of the above equation as
\begin{equation}\label{2.B.47}
\xi^2(t)=\mathcal{B}_1\cos2\Omega t+\mathcal{B}_2\sin 2\Omega t+\frac{\Theta^2+2C|\alpha_0\Lambda_{eg}|^2}{4\Omega^2}
\end{equation}
which has striking similarities with eq.(\ref{2.A.21}). Now $\delta_1^\lambda=\xi^2(t)$ and from eq.(\ref{2.B.41}) with the initial conditions for $\gamma_1^\lambda$ and $\gamma_2^\lambda$, we find out that $\dot{\delta}_1^\lambda(0)=0$. Finally, we can write down the analytical expression for $\xi^2(t)$ as
\begin{equation}\label{2.B.48}
\xi^2(t)=\frac{(2-C)|\alpha_0\Lambda_{eg}|^2}{2\Omega^2}\cos 2\Omega t+\frac{\Theta^2+2C|\alpha_0\Lambda_{eg}|^2}{4\Omega^2}~.
\end{equation}
Similarly, the other coefficient $\gamma_1^\lambda$ reads
\begin{equation}\label{2.B.49}
\begin{split}
\gamma_1^\lambda(t)&=\frac{(E_e-E_g-\omega_0)}{2\alpha_0\Lambda_{eg}}e^{i\omega_0 t}(\xi^2(t)-1)-\frac{i\xi(t)\dot{\xi}(t)e^{i\omega_0 t}}{\alpha_0 \Lambda_{eg}}
\end{split}
\end{equation}
and $\gamma_2^\lambda(t)=\gamma_1^{\lambda^*}(t)$ as $\hat{\mathcal{I}}(t)$ is a Hermitian operator. The only discrepancy lies in the determination of the constant $C$. If $\hat{\mathcal{I}}(t)$ can be expressed as an outer product of a pure state vector $|\phi_I(t)\rangle$ then from the eigen-equation of the scaled invariant operator, we find out that $|\phi_I(t)\rangle=|\phi_{LR}\rangle$. Hence, the initial condition of the state vector $|\phi_{LR}\rangle$ gives $C$ to be unity. For this choice of the constant, $\mathcal{I}(t)$ is identical to the density matrix obtained in the earlier subsection. In the current analysis, the time derivative of the Lewis phase then reads
\begin{equation}\label{2.B.50}
\dot{\theta}(t)=\langle \phi_{LR}(t)|i\frac{\partial}{\partial t}-\hat{H}_R(t)|\phi_{LR}(t)\rangle=\frac{1}{2}(\omega_0-E_e-E_g)~.
\end{equation}
Hence, the Lewis phase is obtained as
\begin{equation}\label{2.B.51}
\theta(t)=\frac{t}{2}(\omega_0-E_e-E_g)
\end{equation}
where the constant of integration is set to zero using the initial condition $\theta(0)=0$. This is a very important observation in our work. We find out that the rescaled invariant operator $\hat{\mathcal{I}}(t)$ is identical to the density matrix of the system, and solving the differential equations corresponding to the coefficients is identical to solving the Liouville equation involving the density matrix of the system. Finally, one important observation is that for the Hamiltonian with periodicity, the Lewis phase is linearly dependent on time, where the time derivative gives the overall phase for the state vector. Hence, Liouville equations are a special case of the equation of motion involving the different classes of invariant operators. With this interesting identification, we move towards solving a two-level system influenced by a time-dependent magnetic field. 
\section{Two-level system influenced by a time-dependent magnetic field}\label{S3}
\noindent We consider a two-level system which is influenced by a constant magnetic field in the $z$-direction and a time-dependent magnetic field in the $x$-direction. The Hamiltonian for such a model is given as
\begin{equation}\label{3.52}
\begin{split}
H(t)&=-\mu B_z (\sigma_z+f(t)\sigma_x)\\
&=-E_0(\sigma_z+f(t)\sigma_x)
\end{split}
\end{equation}
where $\mu B_z=E_0$ and 
\begin{equation}\label{3.53}
f(t)=\left\lbrace{\begin{matrix}
f_0~~t<\frac{T}{2}\\-f_0~~t\geq\frac{T}{2}
\end{matrix}}\right.
\end{equation}
with $f_0\in \mathbb{R}^+$. The periodic square pulse $f(t)$ in the above equation can be written using the Heaviside theta function as
\begin{equation}\label{3.54}
\begin{split}
f(t)=&f_0\Theta\left(\frac{T}{2}-t\right)-f_0\Theta\left(t-\frac{T}{2}\right)\\&-4f_0\Theta\left(\frac{T}{2}-t\right)\Theta\left(t-\frac{T}{2}\right)
\end{split}
\end{equation}
where the Heaviside theta function $\Theta\left(t-\frac{T}{2}\right)$ is defined as
\begin{equation}
\Theta\left(t-\frac{T}{2}\right)=\left\lbrace\begin{matrix}
1&&t>\frac{T}{2}\\
0&&t<\frac{T}{2}\\
\frac{1}{2}&&t=\frac{T}{2}
\end{matrix}\right.~.
\end{equation}
 We consider the density matrix of the system as
\begin{equation}\label{3.55}
\rho(t)=\begin{pmatrix}
\rho_{00}(t)&&\rho_{01}(t)\\
\rho_{10}(t)&&\rho_{11}(t)
\end{pmatrix}~.
\end{equation}
As before, one can obtain four Liouville equations of motion as
\begin{align}
\dot{\rho}_{00}(t)&=-if(t)E_0(\rho_{01}(t)-\rho_{10}(t))\label{3.56}\\
\dot{\rho}_{11}(t)&=if(t)E_0(\rho_{01}(t)-\rho_{10}(t))\label{3.57}\\
\dot{\rho}_{01}(t)&=-if(t)E_0(\rho_{00}(t)-\rho_{11}(t))+2iE_0\rho_{01}(t)\label{3.58}\\
\dot{\rho}_{10}(t)&=if(t)E_0(\rho_{00}(t)-\rho_{11}(t))-2iE_0\rho_{10}(t)\label{3.59}~.
\end{align}
As before, from the first two equations, we obtain $\rho_{00}=1-\rho_{11}$. We again consider the state to be purely in the ground state such that $\rho_{00}(0)=1$ and $\rho_{01}(0)=\rho_{10}(0)=\rho_{11}(0)=0$. Multiplying  eq.(s)(\ref{3.58},\ref{3.59}) with $f(t)$ and summing them,  we obtain a new equation of the form
\begin{equation}\label{3.60}
\begin{split}
f(t)\left(\dot{\rho}_{01}(t)+\dot{\rho}_{10}(t)\right)=&2i f(t) E_0(\rho_{01}(t)-\rho_{10}(t))\\
=&-2\dot{\rho}_{00}(t)\\
\implies \dot{\rho}_{00}(t)=&-\frac{f(t)}{2}\left(\dot{\rho}_{01}(t)+\dot{\rho}_{10}(t)\right)\\
=&-\frac{1}{2}\frac{d}{dt}\left(f(t)(\rho_{01}(t)+\rho_{10}(t))\right)\\
&+\frac{\dot{f}(t)}{2}(\rho_{01}(t)+\rho_{10}(t))
\end{split}
\end{equation}
where in the second line of the above equation, we have made use of eq.(\ref{3.56}). From eq.(\ref{3.54}), we can calculate the analytical form of the time derivative of the function $f(t)$ as
\begin{equation}\label{3.60a}
\begin{split}
\dot{f}(t)=&-2f_0\delta\left(\frac{T}{2}-t\right)-4f_0\biggr[\Theta\left(\frac{T}{2}-t\right)\delta\left(t-\frac{T}{2}\right)\\&-\delta\left(t-\frac{T}{2}\right)\Theta\left(t-\frac{T}{2}\right)\biggr]\\
=&-2f_0\delta\left(t-\frac{T}{2}\right)-2f_0\left[\delta\left(t-\frac{T}{2}\right)-\delta\left(t-\frac{T}{2}\right)\right]\\
=&-2f_0\delta\left(t-\frac{T}{2}\right)~.
\end{split}
\end{equation}
\noindent Substituting the expression of $\dot{f}(t)$ in the last line of eq.(\ref{3.60}) and executing the time integral from $0$ to $t$, we obtain
\begin{equation}\label{3.61}
\begin{split}
\rho_{00}(t)=&\rho_{00}(0)-\frac{f(t)}{2}\left(\rho_{01}(t)+\rho_{10}(t)\right)\\&-f_0\Theta\left(t-\frac{T}{2}\right)\left(\rho_{01}(T/2)+\rho_{10}(T/2)\right)~.
\end{split}
\end{equation}
At $t=T/2$, we can recast the above equation using the analytical form of $f(t)$ in eq.(\ref{3.54}) as
\begin{equation}\label{3.62}
\begin{split}
\rho_{00}(T/2)=&1+\frac{f_0}{2}\left(\rho_{01}(T/2)+\rho_{10}(T/2)\right)\\&-f_0\Theta(0)\left(\rho_{01}(T/2)+\rho_{10}(T/2)\right)
\end{split}
\end{equation}
where we have made use of the initial condition $\rho_{00}(0)=1$. Now using the value of the Heaviside-theta function at $t=T/2$, $\Theta(0)=\frac{1}{2}$, we obtain
\begin{equation}\label{3.63}
\rho_{00}(T/2)=1~.
\end{equation}
As the trace of the density matrix is always unity, we obtain $\rho_{11}(T/2)=0$. We now make a claim that at $t=T/2$ the state remains pure and as a result $\rho^2=\rho$. A careful calculation reveals that in such a scenario $|\rho_{01}(T/2)|=0$. We can therefore write down the following condition at $t=T/2$ for the density matrix as $\rho_{00}(T/2)=1$ and $\rho_{11}(T/2)=\rho_{01}(T/2)=\rho_{10}(T/2)=0$. Using these values back in eq.(\ref{3.61}), we obtain
\begin{equation}\label{3.64}
\begin{split}
\rho_{00}(t)&=1-\frac{f(t)}{2}\left(\rho_{01}(t)+\rho_{10}(t)\right)\\
\implies 2(1-\rho_{00}(t))&=f(t)\left(\rho_{01}(t)+\rho_{10}(t)\right).
\end{split}
\end{equation}
Subtracting eq.(\ref{3.59}) from eq.(\ref{3.58}) and multiplying both sides by $-if(t)E_0$, we obtain a new equation of the form
\begin{equation}\label{3.65}
\begin{split}
-if(t) E_0\left(\dot{\rho}_{01}(t)-\dot{\rho}_{10}(t)\right)=&-2f^2(t) E_0^2(\rho_{00}(t)-\rho_{11}(t))\\&+2f(t) E_0^2(\rho_{01}(t)+\rho_{10}(t))~.
\end{split}
\end{equation}
The left-hand side of the above equation can be expressed as
\begin{equation}\label{3.66}
\begin{split}
-if E_0\left(\dot{\rho}_{01}(t)-\dot{\rho}_{10}(t)\right)&=\frac{d}{dt}\left[-if E_0\left(\rho_{01}(t)-\rho_{10}(t)\right)\right]\\
&+i\dot{f}E_0\left(\rho_{01}(t)-\rho_{10}(t)\right)\\
&=\ddot{\rho}_{00}(t)-\frac{i\dot{f}(t)}{f(t)}\dot{\rho}_{00}(t)\\
&=\ddot{\rho}_{00}(t)+\frac{2i f_0\delta\left[t-\frac{T}{2}\right]}{f\left(\frac{T}{2}\right)}\dot{\rho}_{00}\left[\frac{T}{2}\right]\\
&=\ddot{\rho}_{00}(t)-2i~\delta\left[t-\frac{T}{2}\right]\dot{\rho}_{00}\left[\frac{T}{2}\right]
\end{split}
\end{equation}
where we have made use of eq.(s)(\ref{3.56}) in the second line of the above equation and eq.(\ref{3.54}) in the final two lines of the above equation. Using the values of the off-diagonal elements of the density matrix at $t=\frac{T}{2}$, we obtain from eq.(\ref{3.54}), $\dot{\rho}_{00}\left(\frac{T}{2}\right)=0$. Substituting this value back in eq.(\ref{3.66}), we obtain 
\begin{equation}\label{3.67}
-if(t) E_0\left(\dot{\rho}_{01}(t)-\dot{\rho}_{10}(t)\right)=\ddot{\rho}_{00}(t)~.
\end{equation}
Making use of the above expression along with eq.(\ref{3.64}) in eq.(\ref{3.65}), we obtain a second order differential equation in $\rho_{00}(t)$ as
\begin{equation}\label{3.68}
\ddot{\rho}_{00}(t)+4E_0^2(1+f^2(t))\rho_{00}(t)=2E_0^2(2+f^2(t))~.
\end{equation}
The above equation is very elegant in the sense that for $t<\frac{T}{2}$ and $t>\frac{T}{2}$, eq.(\ref{3.68}) reduces to the same structure given as
\begin{equation}\label{3.69}
\ddot{\rho}_{00}(t)+4E_0^2(1+f_0^2)\rho_{00}(t)=2E_0^2(2+f_0^2)~.
\end{equation}
For $t<\frac{T}{2}$, we obtain a solution for the above equation as
\begin{equation}\label{3.70}
\rho_{00}(t)=\mathcal{A}_{f_0}\cos\left(2\mathcal{E}_0 t\right)+\mathcal{A}'_{f_0}\sin\left(2\mathcal{E}_0 t\right)+\frac{2+f_0^2}{2(1+f_0^2)}
\end{equation}
where $\mathcal{E}_0\equiv E_0 \sqrt{1+f_0^2}$. At $t=0$, $\rho_{00}(0)=1$ and $\dot{\rho}_{00}=0$. These initial conditions help us to completely determine the analytical form of $\rho_{00}(t)$ for $t<\frac{T}{2}$ as
\begin{equation}\label{3.71}
\rho_{00}(t)=\frac{f_0^2}{2(1+f_0^2)}\cos(2\mathcal{E}_0t)+\frac{2+f_0^2}{2(1+f_0^2)}~.
\end{equation}
Similarly for $t>\frac{T}{2}$, and making use of the initial conditions $\rho_{00}(T/2)=1$ and $\dot{\rho}_{00}(T/2)=0$, one can obtain the solution of the equation in eq.(\ref{3.69}) as
\begin{equation}\label{3.72}
\rho_{00}(t)=\frac{f_0^2}{2(1+f_0^2)}\cos(2\mathcal{E}_0t)+\frac{2+f_0^2}{2(1+f_0^2)}~.
\end{equation}
Now, substituting $t=\frac{T}{2}$ in the above equation and also using $\rho_{00}\left(\frac{T}{2}\right)$, we obtain the condition for periodicity to be
\begin{equation}\label{3.73}
T=\frac{2N\pi}{E_0\sqrt{1+f_0^2}}
\end{equation} 
where $N\in\mathbb{Z}^+$. As $\rho_{11}(t)$ is related to $\rho_{00}(t)$ via the relation $\rho_{11}(t)=1-\rho_{00}(t)$, we can obtain the analytical form of the density matrix at all times as
\begin{align}
\rho_{00}(t)&=\frac{f_0^2}{2(1+f_0^2)}\cos(2E_0t\sqrt{1+f_0^2})+\frac{2+f_0^2}{2(1+f_0^2)}\label{3.74}\\
\rho_{11}(t)&=\frac{f_0^2}{1+f_0^2}\sin^2(E_0t \sqrt{1+f_0^2})~.\label{3.75}
\end{align}
We now need to obtain the off-diagonal elements of the density matrix for which we need to make use of eq.(s)(\ref{3.56},\ref{3.64}). For $t<\frac{T}{2}$ and using the solution of $\rho_{00}(t)$ from eq.(\ref{3.74}), we can recast eq.(s)(\ref{3.56},\ref{3.64}) as
\begin{align}
\rho_{01}(t)-\rho_{10}(t)&=-\frac{if_0}{\sqrt{1+f_0^2}}\sin\left(2E_0 t\sqrt{1+f_0^2}\right)\label{3.76}\\
\rho_{01}(t)+\rho_{10}(t)&=\frac{f_0}{1+f_0^2}\left(1-\cos\left(2E_0 t\sqrt{1+f_0^2}\right)\right)~.\label{3.77}
\end{align}
For $t<T/2$, we can then obtain the off-diagonal elements of the density matrix as
\begin{widetext}
\begin{align}
\rho_{01}(t)&=\frac{f_0}{2(1+f_0^2)}\left(1-\cos\left(2E_0 t\sqrt{1+f_0^2}\right)\right)-\frac{if_0}{2\sqrt{1+f_0^2}}\sin\left(2E_0 t\sqrt{1+f_0^2}\right)\label{3.78}\\
\rho_{10}(t)&=\frac{f_0}{2(1+f_0^2)}\left(1-\cos\left(2E_0 t\sqrt{1+f_0^2}\right)\right)+\frac{if_0}{2\sqrt{1+f_0^2}}\sin\left(2E_0 t\sqrt{1+f_0^2}\right)\label{3.79}~.
\end{align}
Similarly for $t>\frac{T}{2}$
\begin{align}
\rho_{01}(t)&=-\frac{f_0}{2(1+f_0^2)}\left(1-\cos\left(2E_0 t\sqrt{1+f_0^2}\right)\right)+\frac{if_0}{2\sqrt{1+f_0^2}}\sin\left(2E_0 t\sqrt{1+f_0^2}\right)\label{3.80}\\
\rho_{10}(t)&=-\frac{f_0}{2(1+f_0^2)}\left(1-\cos\left(2E_0 t\sqrt{1+f_0^2}\right)\right)-\frac{if_0}{2\sqrt{1+f_0^2}}\sin\left(2E_0 t\sqrt{1+f_0^2}\right)\label{3.81}~.
\end{align}
\end{widetext}
Again, using the condition $\rho_{01}(T/2)=\rho_{10}(T/2)=0$ we get back the same periodicity condition obtained earlier in eq.(\ref{3.73}). One important thing to observe from eq.(s)(\ref{3.78},\ref{3.79}) and eq.(s)(\ref{3.80},\ref{3.81}) is that the off-diagonal elements change their respective sign when the time crosses $t=\frac{T}{2}$. This is solely due to the form of the square pulse used in eq.(\ref{3.53}) where $f(t)$ goes from $f_0$ for $t<T/2$ to $-f_0$ for $t\geq \frac{T}{2}$. The periodicity condition implies that at $t=\frac{T}{2},T,\frac{3T}{2},\cdots$ the density matrix represents a pure state $\rho=\begin{pmatrix}
1&&0\\
0&&0
\end{pmatrix}$. In order to extract the state of the system, we take a similar ansatz as before
\begin{equation}\label{3.82}
|\phi\rangle=\kappa_0(t)|0\rangle+e^{i\zeta' t}\kappa_1(t)|1\rangle~.
\end{equation} 
Executing $|\phi\rangle\langle\phi|$ and comparing it with the form of the density matrix, we obtain the state of the system in the two time regimes as
\begin{widetext}
\begin{align}
|\phi_{t<T/2}\rangle&=\left(\cos\left[E_0 t\sqrt{f_0^2+1}\right]+\frac{i}{\sqrt{f_0^2+1}}\sin\left[E_0t\sqrt{f_0^2+1}\right]\right)|0\rangle+\frac{if_0}{\sqrt{f_0^2+1}}\sin\left[E_0t\sqrt{f_0^2+1}\right]|1\rangle\label{3.83}\\
|\phi_{t>T/2}\rangle&=\left(\cos\left[E_0 t\sqrt{f_0^2+1}\right]+\frac{i}{\sqrt{f_0^2+1}}\sin\left[E_0t\sqrt{f_0^2+1}\right]\right)|0\rangle-\frac{if_0}{\sqrt{f_0^2+1}}\sin\left[E_0t\sqrt{f_0^2+1}\right]|1\rangle~.\label{3.84}
\end{align}
\end{widetext}
In subsection (\ref{IIB}), we have already identified the Lewis Invariant operator with the density matrix of the system and as a result, we can calculate the time derivative of the Lewis phase just by calculating $\langle\phi|i\frac{\partial}{\partial}-\hat{H}|\phi\rangle$.
In the regime $t<\frac{T}{2}$, we obtain using eq.(\ref{3.83})
\begin{equation}\label{3.85}
\begin{split}
\langle\phi_{t<T/2}|i\frac{\partial}{\partial t}-\hat{H}|\phi_{t<T/2}\rangle&=0\\
\implies \dot{\theta}_{t<T/2}(t)&=0\\
\implies \theta_{t<T/2}(t)&=\theta_0
\end{split}
\end{equation}
where $\theta_0$ is a time-independent constant. One can set this constant phase to zero for simplicity of analysis. Similarly for $t>\frac{T}{2}$, we again obtain $\langle\phi_{t>T/2}|i\frac{\partial}{\partial}-\hat{H}|\phi_{t>T/2}\rangle=0$ which gives the Lewis phase to be constant as well. This implies that the state of the system is identical to $|\phi_{t<T/2}\rangle$ and $|\phi_{t>T/2}\rangle$ up to some constant phase term.
\subsection{Coherence measures for the density matrix} 
\subsubsection{$l_1$-norm measure of coherence}
\begin{figure}
\begin{center}
\includegraphics[scale=0.22]{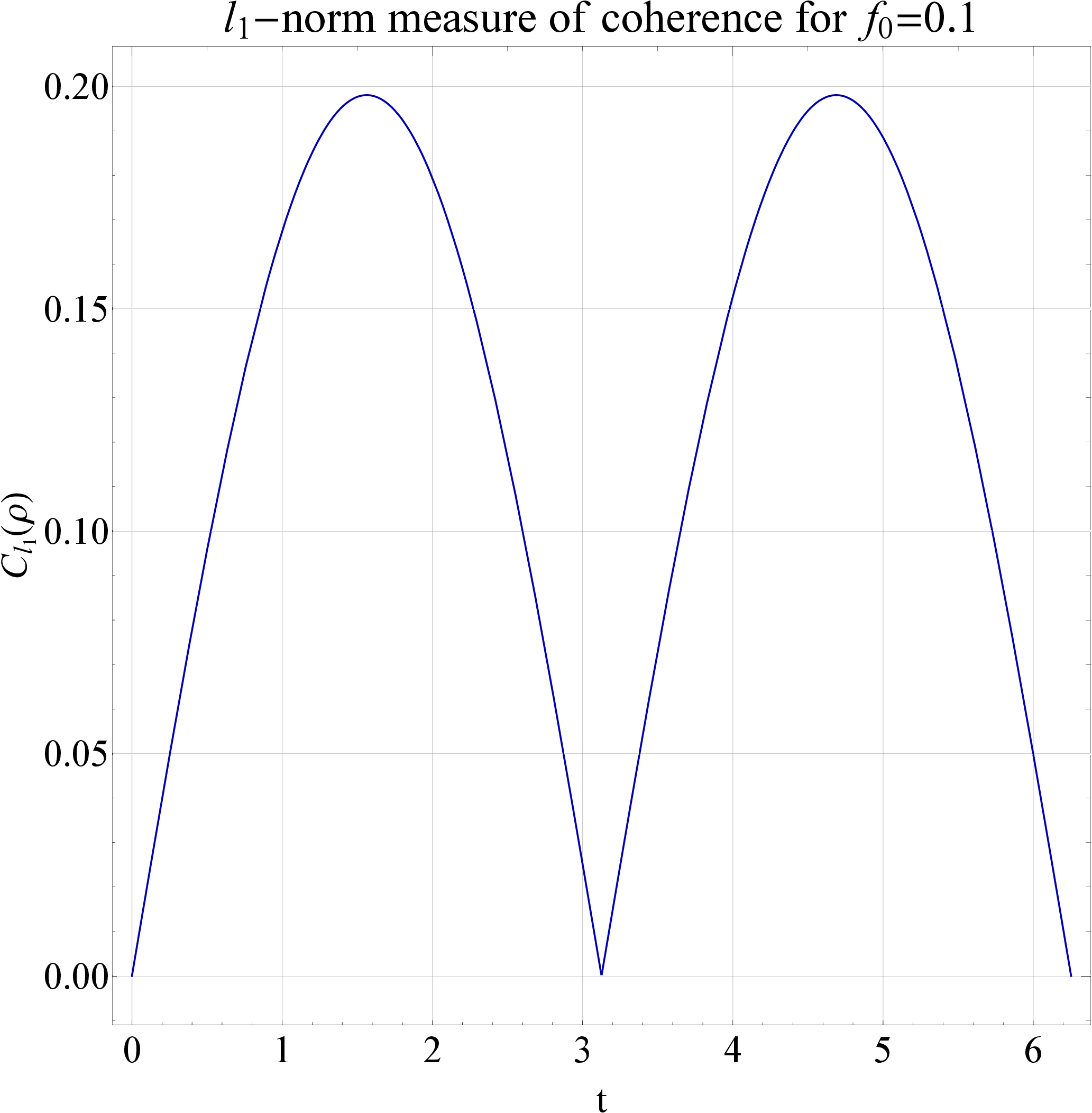}
\caption{$l_1$-norm measure of coherence is plotted against time for $f_0=0.1$.\label{Fig1}}
\end{center}
\end{figure}

\begin{figure}
\begin{center}
\includegraphics[scale=0.22]{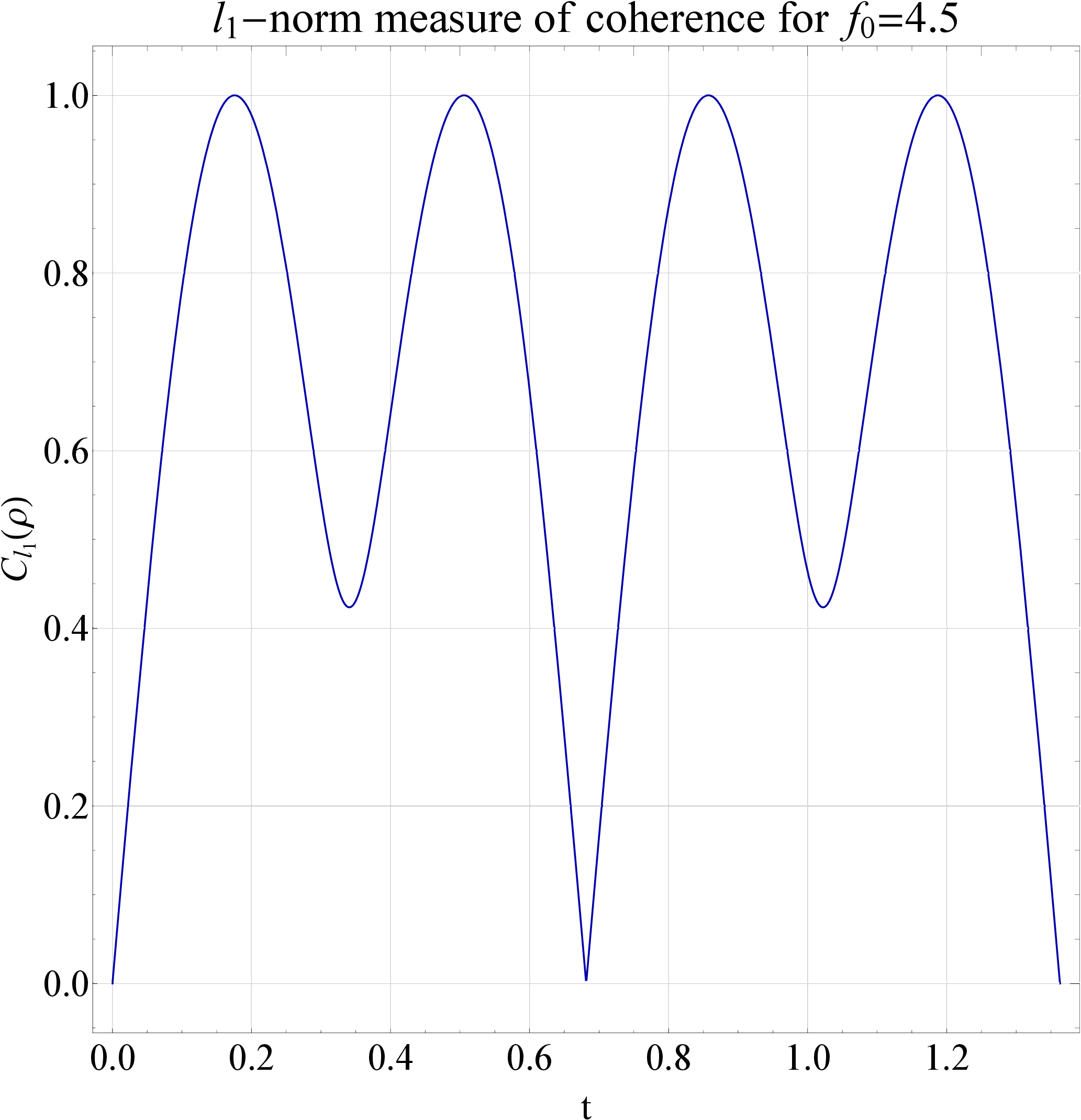}
\caption{$l_1$-norm measure of coherence is plotted against time for $f_0=4.5$.\label{Fig2}}
\end{center} 
\end{figure}

\noindent The $l_1$-norm measure of coherence is defined as \cite{CoherenceMeasure}
\begin{equation}\label{3.A.82}
C_{l_1}(\rho)=\sum\limits_{\substack{{i,j=1}\\,{i\neq j}}}^{d}|\rho_{ij}|~.
\end{equation}
In our model system, $d=2$ and in our current analysis, we obtain the $l_1$-norm measure of coherence as
\begin{equation}\label{3.A.83}
\begin{split}
C_{l_1}(\rho)=\frac{2f_0}{1+f_0^2}\sqrt{\sin^2(\mathcal{E}_0t)\left(1+f_0^2\cos^2(\mathcal{E}_0t)\right)}~.
\end{split}
\end{equation}
We shall now plot the coherence measure with time in the range $0\leq t\leq \frac{2\pi}{E_0\sqrt{1+f_0^2}}$ for $f_0=0.1$ in Fig.(\ref{Fig1}).
We observe from Fig.(\ref{Fig1}) that in the first cycle at the starting time and at $t=T/2$, the measure of coherence is zero, whereas it becomes maximum at $t=T/4$ when $f_0=0.1$. If $f_0$ is greater than unity, then we observe a more interesting behaviour of the coherence measure as can be seen from Fig.(\ref{Fig2}). We observe from Fig.(\ref{Fig2}) that the measure of coherence increases, then drops at $t=T/4$ and again increases and goes to zero at $t=\frac{T}{2}$. This same pattern follows for $\frac{T}{2}\leq t\leq T$. We also observe that the $l_1$-norm measure of coherence becomes unity for $f_0\geq1$ as can be observed from Fig.(\ref{Fig2}). As can be seen from the expression in eq.(\ref{3.A.83}), the coherence among the two states of the two-level system becomes zero provided $f_0=0$, indicating the time-dependent part of the Hamiltonian is responsible for generating this time-dependent coherence between the two states. This result is a bit counter-intuitive as from eq.(s)(\ref{3.83},\ref{3.84}), we already find out that the density matrix is indeed separable for all values of $t$ and as a result there is no true mixing of states. To investigate this issue, we calculate the Frobenius norm measure of coherence.
\subsubsection{Frobenius-norm measure of coherence}
\noindent We shall now proceed towards calculating the Frobenius-norm measure of coherence for the density matrix obtained in section (\ref{S3}). The Frobenius-norm measure of coherence is given by \cite{FrobeniusMeasure}
\begin{equation}\label{3.A.84}
\begin{split}
C_F(\rho)=\sqrt{\frac{d}{d-1}\sum\limits_{i=1}^d\left(\lambda_i-\frac{1}{d}\right)^2}
\end{split}
\end{equation}
where $\lambda_i$ denotes the eigenvalues of the density matrix. In our current analysis $d=2$, and the Frobenius-norm measure of coherence reduces to the standard form
\begin{equation}\label{3.A.85}
C_F(\rho)=\sqrt{2\left(\left(\lambda_1-\frac{1}{2}\right)^2+\left(\lambda_2-\frac{1}{2}\right)^2\right)}~.
\end{equation}
For a density matrix, the eigenvalues read
\begin{equation}\label{3.A.86}
\lambda_{\pm}=\frac{1}{2}\pm\sqrt{\frac{1}{4}+|\rho_{01}|^2-\rho_{00}\rho_{11}}
\end{equation}
where $\lambda_1=\lambda_+$ and $\lambda_2=\lambda_-$. Substituting the eigenvalues in eq.(\ref{3.A.85}), we obtain the analytical form of the Frobenius-norm measure of coherence as
\begin{equation}\label{3.A.87}
C_F(\rho)=\sqrt{1+4|\rho_{01}|^2-4\rho_{00}\rho_{11}}~.
\end{equation}
For both $t<T/2$ and $t>T/2$, $|\rho_{01}|^2$ has the same analytical structure as can be noted by observing the analytical forms of the off-diagonal elements of the density matrix in these two time sectors from eq.(s)(\ref{3.78},\ref{3.79},\ref{3.80},\ref{3.81}). It is easy to check that $|\rho_{01}|^2-\rho_{00}\rho_{11}=0$ from the analytical form of the density matrix. Hence, the Frobenius-norm measure of coherence is
\begin{equation}\label{3.A.88}
C_F(\rho)=1~.
\end{equation}
The above result is indeed very interesting. The unity value of the Frobenius norm measure indicates that the system has geometrically the maximum distance from the maximally mixed state. Hence, the state is always coherent from a geometrical point of view. The primary observation lies in the fact that the $l_1$-norm measure of coherence is time-dependent, whereas the Frobenius-norm measure of coherence is time-independent. This implies that the states are fully coherent from a geometric point of view, which also coincides with the outcome that the density matrix is separable for all values of $t$. Here, $l_1$-norm coherence measure indeed does not capture the full picture as the state is completely separable at $t=0,T/2,\frac{3T}{2}\cdots$, indicating that it is very far from the maximally mixed state, which is the two-dimensional identity matrix with appropriate numerical pre-factor. Hence, for such a scenario, the Frobenius-norm measure of coherence is more reliable than the $l_1$-norm measure of coherence.
\section{Conclusion}\label{S4}
\noindent In this work, we have considered the interaction of two-level systems with time-dependent Hamiltonians which are periodic in time. We resorted to a density matrix approach, where our primary aim was to solve the von Neumann equations of motion to obtain the analytical form of the density matrix. At first, we solved exactly the standard Rabi-oscillation model by solving the four independent differential equations corresponding to the four components of the density matrix. Making use of appropriate initial conditions and the Floquet theorem, we then obtain the state of the system. Next, we have made use of the Lewis-Riesenfeld invariant approach to solve the same model. The interesting observation lies in the fact that the Lewis invariant operator has the same analytical structure as that of the density matrix solved in the initial analysis. This gives us a very robust connection between the density matrix as well as the Lewis invariant approach. We therefore conclude that the density matrix of a system is the same as the Lewis invariant operator for a system up to some constant factor, and we also find out that the Lewis phase of the system for a Hamiltonian with periodicity is some constant multiplied by time. Finally, using this density matrix approach, we went on to 
solve a two-level system influenced by a time-independent constant magnetic field in the $z$-direction and a time-dependent magnetic field in the $x$-direction. For a periodic square pulse, we observe a very interesting solution of the density matrix where the diagonal elements remain the same over time and the off-diagonal elements pick up a negative sign when $t$ crosses $T/2$. From the value of the elements of the density matrix at $t=T/2$, we obtain a periodicity condition. We find out that the density matrix is separable for all times for this two-level system influenced by a periodic magnetic field, and the state has a different structure for $t<T/2$ 
and $t>T/2$. We also find out that for $t>T/2$, the coefficient corresponding to state $|1\rangle$ changes its sign. We have then plotted the $l_1$-norm measure of coherence, which depicts an explicit time dependence, and the coherence becomes maximum ($C_{l_1}(\rho)=1$) for $f_0\geq 1$. On the other hand, if we compute the Frobenius-norm measure of coherence, we find that the measure of coherence is unity and independent of time, indicating a fully coherent state throughout, which is more reliable, as we have found that the density matrix is separable for all values of $t$. Hence, it is interesting to find out that for such a scenario, the Frobenius-norm measure of coherence, which is a geometric interpretation of the distance of a state from the maximally entangled state, depicts the true nature of coherence, implying that the $l_1$-norm measure of coherence may not be reliable always.
\section*{Acknowledgement}
\noindent We thank the anonymous referee for the useful comments that have helped to improve our paper.
\section*{Data availability statement}
\noindent This manuscript has no associated data, and no new data were generated during the completion of the manuscript.


\begin{thebibliography}{8}
\bibitem{Landau}
L. D. Landau, ``\textit{On the Theory of Transfer of Energy at Collisions II}", Phys. Z. Sowjetunion 2 (1932) 46.
\bibitem{Zener}
C. Zener, ``\textit{Non-adiabatic crossing of energy levels}", \href{http://dx.doi.org/10.1098/rspa.1932.0165}{Proc. R. Soc. A 137 (1932) 696}.
\bibitem{Rabi}
I. I. Rabi, ``\textit{Space Quantization in a Gyrating Magnetic Field}", \href{https://journals.aps.org/pr/abstract/10.1103/PhysRev.51.652}{Phys. Rev. 51 (1937) 652}.
\bibitem{BarnsSarma}
E. Barns and S. D. Sarma, ``\textit{Analytically Solvable Driven Time-Dependent Two-Level Quantum Systems}", \href{https://journals.aps.org/prl/abstract/10.1103/PhysRevLett.109.060401}{Phys. Rev. Lett. 109 (2012) 060401}.
\bibitem{Rydberg}
Y. O. Duding, L. Li, F. Bariani, and A. Kuzmich, ``\textit{Observation of coherent many-body Rabi oscillations}", \href{https://www.nature.com/articles/nphys2413}{Nat. Phys. 12 (2012) 790}. 
\bibitem{Atomic1}
P. L. Knight and P. W. Milonni, ``\textit{The Rabi frequency in optical spectra}", \href{https://doi.org/10.1016/0370-1573(80)90119-2}{Phys. Rep. 66 (1980) 21}.
\bibitem{Atomic2}
B. W. Shore and P. L. Knight, ``\textit{The Jaynes-Cunning model}", \href{https://www.tandfonline.com/doi/abs/10.1080/09500349314551321}{J. Mod. Opt. 40 (1993) 1195}.
\bibitem{Josephson}
J. M. Martinis, S. Nam, J. Aumentado, and C. Urbina, ``\textit{Rabi oscillations in a large Josephson-junction qubit}", \href{https://doi.org/10.1103/PhysRevLett.89.117901}{Phys. Rev. Lett. 89 (2002) 117901}.
\bibitem{Josephson1}
N. Gr\o{}nbech-Jensen and M. Cirillo, ``\textit{Rabi-Type Oscillations in a Classical Josephson Junction}", \href{https://doi.org/10.1103/PhysRevLett.95.067001}{Phys. Rev. Lett. 95 (2005) 067001}.
\bibitem{Josephson2}
A. Rahmani and F. B. Laussy, ``\textit{Polaritonic Rabi and Josephson Oscillations}", \href{https://www.nature.com/articles/srep28930}{Sci. Rep. 6 (2016) 28930}.
\bibitem{Cavity1}
G. S. Agarwal, ``\textit{Vacuum-field Rabi oscillations of atoms in a cavity}", \href{https://doi.org/10.1364/JOSAB.2.000480}{J. Opt. Soc. America B 2 (1985) 480}.
\bibitem{Cavity2}
R. Houdr\'{e}, R. P. Stanley, U. Oesterle, M. Ilegems, and C. Weisbuch, ``\textit{Room-temperature cavity polaritons in a semiconductor microcavity}", \href{https://doi.org/10.1103/PhysRevB.49.16761}{Phys. Rev. B 49 (1994) 16761}.
\bibitem{Cavity3}
H. Deng, H. Haug, and Y. Yamamoto, ``\textit{Exciton-polariton Bose-Einstein condensation}", \href{https://doi.org/10.1103/RevModPhys.82.1489}{Rev. Mod. Phys. 82 (2010) 1489}.
\bibitem{Cavity4}
A. Amo, J. Lefr\`{e}re, S. Pigeon, C. Adrados, C. Ciuti, I. Carusotto, R. Houdr\'{e}, E. Giacobino, and A. Bramati, ``\textit{Superfluidity of polaritons in in semiconductor microcavities}", \href{https://doi.org/10.1038/nphys1364}{Nat. Phys. 5 (2009) 805}.
\bibitem{Cavity5}
P.-O. Guimond, A. Roulet, H. N. Le, and V. Scarani, ``\textit{Rabi oscillation in a quantum cavity: Markovian and non-Markovian dynamics}", \href{https://doi.org/10.1103/PhysRevA.93.023808}{Phys. Rev. A 93 (2016) 023808}.
\bibitem{AJPMerlin}
R. Merlin, ``\textit{Rabi oscillations, Floquet states, Fermi's golden rule, and all insights from an exactly solvable two-level model}", \href{https://doi.org/10.1119/10.0001897}{Am. J. Phys. 89 (2021) 26}.
\bibitem{Floquet1}
G. Floquet, ``\textit{Sur les \'{e}quations diff\'{e}rentielles lin\'{e}aires \`{a} coefficients p\'{e}riodiques}", \href{https://doi.org/10.24033/asens.220}{Ann. Sci. de l'\'{E}cole Norm. Sup\'{e}rieure 12 (1883) 47}.
\bibitem{Floquet2}
J. H. Shirley, ``Solution of the Schr\"{o}dinger Equation with a Hamiltonian Periodic in Time", \href{https://doi.org/10.1103/PhysRev.138.B979}{Phys. Rev. 198 (1965) B979}.
\bibitem{FloquetState0}
T. Oka and H. Aoki, ``\textit{Photovoltaic Hall effect in graphene}", \href{https://doi.org/10.1103/PhysRevB.79.081406}{Phys. Rev. B 79 (2009) 081406(R)}.
\bibitem{FloquetState1}
N. H. Linder, H. Netanel, G. Refael, and V. Galitski, ``\textit{Floquet topological insulator in semiconductor quantum wells}", \href{https://doi.org/10.1038/nphys1926}{Nat. Phys. 7 (2011) 490}.
\bibitem{FloquetState2}
Y. H. Wand, H. Steinberg, P. Jarillo-Herraro, and N. Gedik, ``\textit{Observation of Floquet-Bloch states on the surface of a topological insulator}", \href{https://doi.org/10.1126/science.1239834}{Science 342 (2013) 453}.
\bibitem{FloquetState3}
M. C. Rechtsman, J. M. Zeuner, Y. Plotnik, Y. Lumer, D. Podolsky, F. Dreisow, S. Nolte, M. Segev, and A. Szameit, ``\textit{Photonic Floquet topological insulators}", \href{https://doi.org/10.1038/nature12066}{Nature 496 (2013) 196}.
\bibitem{FloquetState4}
M. A. Sentef, M. Claassen, A. F. Kemper, B. Moritz, T. Oka, J. K. Freericks, and T. P. Devereaux, ``\textit{Theory of Floquet band formation and local pseudospin textures in pump-probe photoemission of graphene}", \href{https://doi.org/10.1038/ncomms8047}{Nat. Commun. 6 (2015) 7047}.
\bibitem{HRLewis1}
H. R. Lewis Jr., ``\textit{Classical and Quantum Systems with Time-Dependent Harmonic-Oscillator-Type Hamiltonians}", \href{https://doi.org/10.1103/PhysRevLett.18.510}{Phys. Rev. Lett. 18 (1967) 510}.
\bibitem{HRLewis2}
H. R. Lewis Jr., ``\textit{Class of Exact Invariants for Classical and Quantum Time-Dependent Harmonic Oscillators}", \href{https://doi.org/10.1063/1.1664532}{J. Math. Phys. 9 (1968) 1976}.
\bibitem{CoherenceMeasure}
T. Baumgratz, M. Cramer, and M. B. Plenio, ``\textit{Quantifying coherence}", \href{https://doi.org/10.1103/PhysRevLett.113.140401}{Phys. Rev. Lett. 113 (2014) 140401}.
\bibitem{FrobeniusMeasure}
Y. Yao, G. H. Dong, X. Xiao, and C. P. Sun, ``\textit{Frobenius norm-based measures of quantum coherence and asymmetry}", \href{https://www.nature.com/articles/srep32010}{Sci. Rep. 6 (2016) 32010}.
\bibitem{LewisRiesenfeld}
H. R. Lewis Jr. and W. B. Riesenfeld, ``\textit{An Exact Quantum Theory of Time-Dependent Harmonic Oscillator and of a Charged Particle in a Time-Dependent Electromagnetic Field}", \href{https://doi.org/10.1063/1.1664991}{J. Math. Phys. 10 (1969) 1458}.
\end{thebibliography}
\end{document}